\documentclass{bmcart}

\usepackage{amsthm,amsmath}
\usepackage{graphicx}
\usepackage{float} 
\RequirePackage{hyperref}
\usepackage[utf8]{inputenc} 
\usepackage{color}

\startlocaldefs

\endlocaldefs

\begin{document}

\begin{frontmatter}

\begin{fmbox}
\dochead{Research}

\title{Numerical Simulations of Noisy Quantum Circuits for Computational Chemistry}


\author[addressref={aff1}, noteref={n1}]{\inits{J.B.}\fnm{Jerimiah}~\snm{Wright}}
\author[addressref={aff3,aff2},email={mgowrish@vols.utk.edu}, corref={aff3},noteref={n1}]{\inits{M.G.}\fnm{Meenambika}~\snm{Gowrishankar}}
\author[addressref={aff4}]{\inits{D.C.}\fnm{Daniel}~\snm{Claudino}}
\author[addressref={aff1}]{\inits{P.C.L.}\fnm{Phillip C.}~\snm{Lotshaw}}
\author[addressref={aff4}]{\inits{T.N.}\fnm{Thien}~\snm{Nguyen}}
\author[addressref={aff2,aff4}]{\inits{A.M.}\fnm{Alexander J.}~\snm{McCaskey}}
\author[addressref={aff1,aff3,aff2},email={humblets@ornl.gov},]{\inits{T.S.H}\fnm{Travis S.}~\snm{Humble}}


\address[id=aff1]{
  \orgdiv{Quantum Computational Sciences Group},             
  \orgname{Oak Ridge National Laboratory},          
  \city{Oak Ridge, Tennessee},                              
  \cny{USA}                                    
}
\address[id=aff3]{%
  \orgdiv{Quantum Science Center},
  \orgname{Oak Ridge National Laboratory},
  \city{Oak Ridge, Tennessee},
  \cny{USA}
}
\address[id=aff2]{%
  \orgdiv{Bredesen Center},
  \orgname{University of Tennessee},
  \city{Knoxville, Tennessee},
  \cny{USA}
}

\address[id=aff4]{%
  \orgdiv{Beyond Moore Computing Group},
  \orgname{Oak Ridge National Laboratory},
  \city{Oak Ridge, Tennessee},
  \cny{USA}
}


\begin{artnotes}
\note{\small{This manuscript has been authored by UT-Battelle, LLC under Contract No. DE-AC05-00OR22725 with the U.S. Department of Energy. The United States Government retains and the publisher, by accepting the article for publication, acknowledges that the United States Government retains a non-exclusive, paid-up, irrevocable, world-wide license to publish or reproduce the published form of this manuscript, or allow others to do so, for United States Government purposes. The Department of Energy will provide public access to these results of federally sponsored research in accordance with the DOE Public Access Plan (http://energy.gov/downloads/doe-public-access-plan).}
}
\note[id=n1]{Equal Contributor}
\end{artnotes}

\end{fmbox}


\begin{abstractbox}
\begin{abstract} 
The opportunities afforded by near-term quantum computers to calculate the ground-state properties of small molecules depend on the structure of the computational ansatz as well as the errors induced by device noise. Here we investigate the behavior of these noisy quantum circuits using numerical simulations to estimate the accuracy and fidelity of the prepared quantum states relative to the ground truth obtained by conventional means. We implement several different types of ansatz circuits derived from unitary coupled cluster theory for the purposes of estimating the ground-state energy of sodium hydride  using the variational quantum eigensolver algorithm. We show how relative error in the energy and the fidelity scale with the levels of gate-based noise, the internuclear configuration, the ansatz circuit depth, and the parameter optimization methods.

\end{abstract}


\begin{keyword}
\kwd{Variational Quantum Algorithms}
\kwd{Noise}
\kwd{Quantum Chemistry}
\kwd{Quantum Computing}
\end{keyword}

\end{abstractbox}
%

\end{frontmatter}


\section{Introduction}
Computation is a staple of predicting the behaviors and properties of new chemicals and materials, and leading computational methods must take advantage of state-of-the-art computers to address the most challenging calculations. The advent of quantum computers has motivated the development and testing of new methods for solving familiar problems by using quantum mechanics itself to perform the computation \cite{whitfield2011simulation,RevModPhys.92.015003,cao2019quantum,bauer2020quantum}. Current quantum computing methods range from fault-tolerant quantum error corrected algorithms that support arbitrarily long calculations to noisy intermediate-scale quantum (NISQ) algorithms that adapt to the error-prone devices available today \cite{preskill2018quantum}. We shall focus on the latter setting but we note the underlying quantum mechanical model affords novel opportunities to simulate the structure and dynamics of chemical systems within both regimes.
\par 
Among several approaches to NISQ algorithms, the family of variational methods have emerged in recent years as especially promising for testing the behavior of noisy quantum devices. Typified by the variational quantum eigensolver (VQE) \cite{mcclean2016theory}, these algorithms use tunable quantum circuits to prepare approximations to the quantum mechanical states of a model Hamiltoninan \cite{PhysRevX.6.031007}. The familiar variational principle offers a promise that the parameters which minimize the observed energy also prepare the best approximation to the Hamiltonian ground state, while modifications of this approach may be used to approximate higher energy eigenstates, i.e., excited states
\cite{claudino2021improving}.
\par 
Multiple experimental investigations have tested the principles of VQE for recovering the electronic states of small molecular Hamiltonians \cite{kandala2017hardware, mccaskey2019quantum,doi:10.1063/1.5141835, yeter2021benchmarking,arute2020hartree}. Those results indicate that the choice of the parameterized quantum circuit, i.e., the ansatz, plays a significant role in the accuracy of the estimated energy as well as the complexity of finding the optimal parameters. In particular, development of an expressive and efficient ansatz is essential to temper the influence of noise and errors that arise during experimental implementation \cite{grimsley2019trotterized,bauman2019quantum,xia2020qubit,metcalf2020resource,mizukami2020orbital,elfving2021simulating,tkachenko2021correlation}. Many different techniques, including symmetry, constraints, and iterative approximations, have been used to further the construction of efficient ansatz circuits \cite{grimsley2019adaptive,gard2020efficient,tang2021qubit}.
\par 
Despite the continuing improvement of such experimental demonstrations, there is an outstanding question as to how the fidelity of the prepared quantum state compares to the idealized pure state representation of the sought electronic ground state. All demonstrations to date are well within the range of conventional computation yet the leading observable for comparison is the energy estimate and not the state. By construction, the applied ansatz circuit has a best approximation to the true ground state of the given Hamiltonian operator,
but the accuracy with which this state is found is not the leading metric for experimental validation. This is due primarily to the complexity of estimating the prepared state, which requires tomographic techniques for reconstructing the quantum state \cite{PhysRevA.100.010302,PhysRevResearch.2.023048}. 
\par
Previously, Claudino et al.~investigated the fidelity of several variational methods in approximating the ground state of few-electron molecular models \cite{claudino2020benchmarking}. Those results show that VQE methods with sophisticated ansatz circuits could approximate the ground electronic state with very high fidelity in the absence of circuit noise, depending on the ansatz and method of parameter selection. However, implementations on NISQ devices are prone to noise that leads to errors in state preparation, and there is an outstanding need to assess how noise influences circuit accuracy. The effects of noise and errors invalidate the pure-state representation used to motivate VQE methods and lead to significant sources of uncertainty in the prepared state.
\par 
Here we examine the accuracy and precision of the ansatz in several VQE methods by using numerical simulations of the noisy quantum circuits. This includes the nominal form of VQE with one- and two-parameter ansatz circuits derived from unitary coupled cluster (UCC) theory as well as the more sophisticated ADAPT-VQE. In both instances, the ansatz is inspired by how coupled cluster theory approximates the ground state based on a basis of singly and doubly excited determinants (UCCSD) as well as a basis of singlet adapted configurations (singlet-adapted UCCSD). We investigate variations in parameter optimization due to noisy gate operations, testing the COBYLA\cite{cobyla} (derivative-free) and BFGS\cite{lbfgs, lbfgs2} (gradient-based) optimizers, and in all cases we consider the influence of gate noise on the estimated energy and prepared state fidelity. As a test case, we simulate VQE calculations of the expected ground state and energy for sodium hydride (NaH) and we compare with conventional electronic structure calculations.
\par 
The remainder of the presentation is organized as follows: in Sec.~\ref{sec:methods}, we present the background on the theory for the VQE methods and ansatz circuits; in Sec.~\ref{sec:Numerical Methods} we present the methods in which we conducted the noisy and noiseless numerical simulations; in Sec.~\ref{sec:results}, we present results from numerical simulations of the noisy ansatz circuits and we make comparisons with conventional solutions in terms of energy and state fidelity; and in Sec.~\ref{sec:conc}, we discuss conclusions from our analysis.
\section{Variational Quantum Eigensolver Methods}
\label{sec:methods}
Within the context of computational chemistry, the non-relativistic, time-independent molecular Hamiltonian is
\begin{equation}
    \hat{H}_\textrm{mol} = -\sum_{i}{\frac{\nabla_{R_i}^2}{2M_i} } -\sum_{i}{\frac{\nabla_{r_i}^2}{2 m_i} } -\sum_{i,j}{\frac{Z_i}{|R_i - r_j|} }
    +\sum_{i,j>i}{\frac{Z_i Z_j}{|R_i - R_j|} } +\sum_{i,j>i}{\frac{1}{|r_i - r_j|} }
\end{equation}
with $R_i$ the coordinate of the $i$-th nuclei of mass $M_i$ and charge $Z_i$ and with $r_i$ the coordinate of the $i$-th electron with mass $m_e = 1$ and charge $e = 1$. Under the Born-Openheimer approximation, the molecular Hamiltonian is decomposed into an electronic Hamiltonian and nuclear Hamiltonian. By adopting a spin-orbital basis $\{\varphi_{p}: p=1,\dots,N \}$ for the electronic degrees of freedom, the second-quantized form of the electronic Hamiltonian is expressed as
\begin{equation}
\label{ref:hferm}
    H(R) = \sum_{pq}{h_{pq}(R) a^{\dagger}_p a_{q}} + \frac{1}{2} \sum_{pqst}{h_{pqst}(R) a^{\dagger}_p a^{\dagger}_q a_{s} a_{t}}
\end{equation}
where the one-electron integrals
\begin{equation}
    h_{pq}(R) = \int{d \sigma \varphi_{p}^{*}(\sigma) \left(\frac{\nabla_{r}^{2}}{2} - \sum_{i}{\frac{Z_{i}}{|R_{i} - r|}} \right) \varphi_{q}(\sigma)}
\end{equation}
and two-electron integrals
\begin{equation}
      h_{pqst} = \int{d \sigma_1 d \sigma_2 \frac{ \varphi_{p}^{*}(\sigma_1) \varphi_{q}^{*}(\sigma_2) \varphi_{s}(\sigma_1) \varphi_{t}(\sigma_2)}{|r_1 - r_2|}
      }  
\end{equation}
are taken with respect to the variable $\sigma$ denoting both spin and position and depend on the nuclear coordinates $R$. The associated fermionic creation and annihilation operators satisfy the anti-commutation relations
\begin{equation}
    \{a^{\dagger}_{p}, a_{q}\} = \delta_{p,q}\hspace{1cm}\textrm{and}\hspace{1cm}\{a_{p}, a_{q}\} = 0
\end{equation}
\par
The fermionic representation of the Hamiltonian in \eqref{ref:hferm} may be transformed into a qubit representation using the usual Pauli operators
\begin{equation}
X = \left(\begin{array}{cc}
        0 & 1 \\
        1 & 0
    \end{array}\right),
\hspace{0.5cm}
Y = \left(\begin{array}{cc}
        0 & -i \\
        i & 0
    \end{array}\right),
\hspace{0.5cm}\textrm{and}\hspace{0.5cm}
Z = \left(\begin{array}{cc}
        1 & 0 \\
        0 & -1
\end{array}\right)
\end{equation}
which satisfy $[X,Y] = -2 i Z$. Several transformations into the qubit representation are known to satisfy the necessary fermionic commutation relations, and we employ the Jordan-Wigner transformation defined as
\begin{equation}
\label{eq:jw}
    a^{\dagger}_{p} = \otimes_{i<p} Z_i \otimes \sigma^{-}_p,\hspace{0.5cm}\textrm{and}\hspace{0.5cm}a_{p} = \otimes_{i<p} Z_i \otimes \sigma^{+}_p
\end{equation}
in terms of the qubit raising and lower operations  $\sigma^{\pm}_{p} = (X_p\pm iY_p)/\sqrt{2}$. The implementation of this transformation requires $n \geq N $ qubits, and the resulting representation of the fermionic Hamiltonian is 
\begin{equation}
\label{eq:hamp}
    H(R) = \sum_{j}{c_{j}(R) P_{j}}
\end{equation}
where $P_j$ denotes the $j$-th string of Pauli operators over $n$ qubits and $c_{j}(R)$ is the corresponding coefficient. The maximum number of terms in \eqref{eq:hamp} is $4^{n}$ but, in practice, symmetries within the Hamiltonian significantly reduce the number of terms that have non-zero coefficients \cite{romero2018strategies}.
\par 
The variational quantum eigensolver (VQE) method estimates the minimal expectation value of a Hermitian operator with respect to a variable quantum circuit. The method relies on the variational principle, which states that only the lowest eigenstate of a non-negative operator can minimize the expectation value. Here, we use the the qubit representation of the electronic Hamiltonian as the operator of interest, such that the estimated energy expectation value is
\begin{equation}
    E(R;\theta^{\star}) = \min_{\theta}{ \langle \psi(\theta)|H(R)|\psi(\theta)\rangle}
\end{equation}
where $|\psi(\theta) \rangle = U(\theta) | \psi(0) \rangle $ is a variable pure quantum state prepared by a (unitary) ansatz operator $U(\theta)$. The parameter $\theta^{\star}$ denotes the optimal value that minimizes the energy, and a  generalization of $\theta$ may include multiple parameters within the ansatz.
\par
The critical choice in applying the VQE method to a  given Hamiltonian is selection of the ansatz operator $U(\theta)$ and the underlying reference state $\psi(0)$. The latter may be selected by using conventional approximations to the electronic ground state that are efficiently encoded as superpositions of binary states. For example, using the Hartree-Fock solution to the electronic Hamiltonian offers a convenient choice for the reference state, and the vacancy or occupation of a molecular orbital may be encoded by 0 or 1, respectively. The choice of the ansatz operator may also be drawn from conventional electronic structure theory, and many recent efforts have focused on unitary coupled cluster (UCC) theory to generate possible choices for the ansatz operator \cite{romero2018strategies,sokolov2020quantum}. The accuracy with which a given ansatz operator represents the true ground state of the Hamiltonian may be quantified using the fidelity
\begin{equation}
    F = |\langle \Psi|U(\theta)|\psi(0) \rangle|^2
\end{equation}
where $\Psi$ is the expected ground state in the qubit representation. We next review a family of ansatz operators derived from UCC theory.
\subsection{Unitary Coupled Cluster Ansatz Operators}
Consider the fermionic UCC ansatz operator \cite{romero2018strategies}
\begin{equation}
\label{eq:uuu}
    U(\theta) = \exp{\left(T(\theta) - T^{\dagger}(\theta)\right)}
\end{equation}
where 
\begin{equation}
    T(\theta) = \sum_{k=1}^{M}{T_{k}(\theta)}
\end{equation}
and $T_{k}(\theta)$ represents the $k$-th cluster operator of $M$ electrons excited. The unitary operator \eqref{eq:uuu} may be approximated to first order by the finite series \cite{PhysRevA.98.022322}
\begin{equation}
    U(\theta) \approx \prod_{m=1}^{M}{e^{\tau_{m}(\theta)} }
\end{equation}
with $\tau_{m}(\theta) = T_{m}(\theta) - T_{m}^{\dagger}(\theta)$ and $M$ is the series limit. Terminating the series at $M=2$ limits the theory to single and double excitations and yields what is known as the UCC singles and doubles (UCCSD) ansatz 
\begin{equation}
\label{eq:usd}
    U_{\textrm{SD}}(\theta) = e^{\tau_{1}(\theta)}  e^{\tau_{2}(\theta)}
\end{equation}
where
\begin{equation}
    T_{1}(\theta) =  \sum\limits_{\mathop{ i\in\textrm{occ}}\limits_{a\in\textrm{virt}}}{\theta_{i}^{a} \hat{t}_{i}^{a} }
\end{equation}
and
\begin{equation}
    T_{2}(\theta) = \sum\limits_{\mathop {i,j \in {\text{occ}}}\limits_{a,b \in {\text{virt}}} } {\theta _{i,j}^{a,b} \hat{t}_{i,j}^{a,b}}
\end{equation}
with $\hat{t}_{i}^{a} = a^{\dagger}_{a} a_{i}$, $\hat{t}_{i,j}^{a,b} = a_a^\dagger a_b^\dagger {a_i}{a_j}$
and the vector $\theta$ defines the single- and double-excitation parameters $\theta^{a}_{i}$ and $\theta^{a,b}_{i,j}$, respectively, for excitations from occupied orbitals $i$ and $j$ to virtual orbitals $a$ and $b$ of the reference state.
\par
The ansatz operator in \eqref{eq:usd} may be transformed by the Jordan-Wigner transformation \eqref{eq:jw} to yield the individual terms
\begin{equation}
    \tau_{1}(\theta) = \frac{i}{2} \sum_{i,a} \theta^{a}_{i} P^{a}_{i}
\end{equation}
and
\begin{equation}
    \tau_{2}(\theta)  = \frac{i}{8}\sum\limits_{\mathop {i,j \in {\text{occ}}}\limits_{a,b \in {\text{virt}}} } 
    {\theta^{a,b}_{i,j} Q_{i,j}^{a,b}     }
\end{equation}
with
\begin{equation}
    P_{i}^{a} = \otimes_{k=i}^{a-1} Z_{k} \left(X_{a} Y_{i} - Y_{a} X_{i} \right)
\end{equation}
and
\begin{equation}
\begin{array}{cc}
    Q_{i,j}^{a,b} =  \otimes_{k=i}^{a-1} Z_{k} \otimes_{\ell = j}^{b-1} Z_{\ell}  & \left( X_a X_b X_i Y_j + X_a Y_b Y_i Y_j + X_a X_b Y_i X_j + Y_a X_b Y_j Y_i \right. \\
    & \left. - Y_a Y_b Y_i X_j  - Y_a X_b X_i X_j - Y_a Y_b X_i Y_j - X_a Y_b X_i X_j   \right)
\end{array}
\end{equation}
Recalling that $X_k = i Y_k Z_k$, the latter may be recast as
\begin{equation}
\label{eq:qijab}
\begin{array}{cc}
    Q_{i,j}^{a,b} =  \otimes_{k=i}^{a-1} Z_{k} \otimes_{\ell = j}^{b-1} Z_{\ell}  & X_a X_b X_i Y_j \left(I_a I_b I_i I_j - I_a Z_b Z_i I_j + I_a I_b Z_i Z_j - Z_a I_b Z_j I_i \right. \\
    & \left. + Z_a Z_b Z_i Z_j  - Z_a I_b I_i Z_j + Z_a Z_b I_i I_j - I_a Z_b I_i Z_j   \right)
\end{array}
\end{equation}
where, under the $Z$ operator, occupied orbitals yield an eigenvalue $+1$ and virtual orbitals yield an eigenvalue $-1$ such that each operator $Q^{a,b}_{i,j}$ is determined by the single four-qubit Pauli string $X_a X_b X_i Y_j$.
\par
The UCCSD operator serves as our starting point for developing various ansatz operators. First, we define a one-parameter ansatz operator derived from the UCCSD ansatz by considering only contribution from the doubles excitations in \eqref{eq:qijab}. For the four-qubit encoding of two-electrons described below, the resulting UCC doubles (UCCD) ansatz operator is reduced to
\begin{equation}
\label{eq:yxxx}
    U(\theta) = e^{i\theta Y_0 X_1X_2X_3}
\end{equation}
We also consider a singlet-adapted variation of the UCCSD ansatz for which the state is defined in a basis of restricted determinants. Specifically, we restrict the ansatz to basis states that are, by construction, spin eigenstates with $S^2=0$. Such states are formed from linear combinations of the corresponding excitation operators that  conserve electron spin and obey the spin symmetry. These linear combinations are often referred to in chemistry as configuration state functions or symmetry-adapted basis functions. We defer the description of constructing these linear combinations to prior work \cite{SzaboOstlund}. Various details on both ansatz operators have been reported previously \cite{mccaskey2019quantum,claudino2020benchmarking}. 
\begin{figure}
    \centering 
    \includegraphics[scale = 0.4]{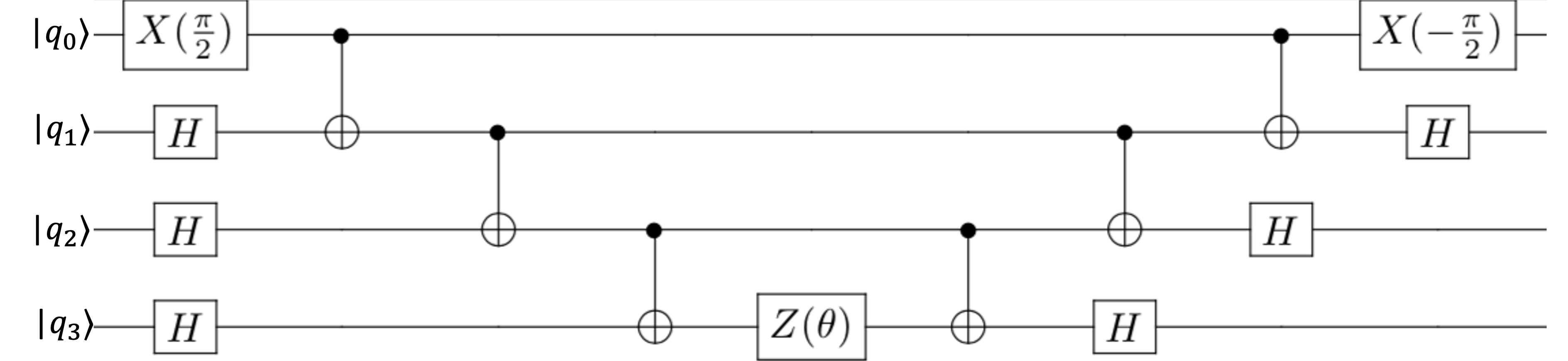}
    \caption{The quantum circuit diagram representing the one-parameter UCCD ansatz used in the numerical studies reported here. This ansatz was originally reported in Ref.~\cite{mccaskey2019quantum}.
    }
    \label{fig:UCC1}
\end{figure}
\par 
An alternative to the fixed-depth ansatz operators above is an adaptive ansatz that changes the generating operator based on iterative evaluations of the energy estimate. Grimsley et al.~first proposed the adaptive derivative-assembled pseudo-trotter (ADAPT) ansatz motivated by UCC theory in which the unitary \eqref{eq:uuu} is expanded beyond first order \cite{grimsley2019adaptive}. They showed how the resulting operator expression guides iteratively growing an ansatz circuit until a desired convergence in the expectation value is obtained. The ADAPT method begins by decomposing the individual single- and double-excitations operators represented by \eqref{eq:usd} into a set known as the operator pool. Let $\mathrm{T} = \{\hat{t}_{i}^{a}\} \bigcup \{\hat{t}_{i,j}^{a,b}\}$ denote the operator pool derived from the UCCSD approximation. An exact ansatz operator
may be expressed as
\begin{equation}
\label{eq:uadapt}
    U_{SD}(\theta) = \prod_{k}^{\infty}{
    \prod_{i,a}{
    \exp{\left[\theta_{i,k}^{a} (\hat{t}_{i}^{a}-\hat{t}_{a}^{i})
    \right]}
    }
    \prod_{i,j,a,b}{ 
    \exp{\left[\theta_{i,j,k}^{a,b} (\hat{t}_{i,j}^{a,b}-\hat{t}_{a,b}^{i,j})
    \right]}
    }
    }
\end{equation}
in which $k$ indexes the different instances of each excitation operator. 
\par 
The ADAPT method creates an approximation to \eqref{eq:uadapt} by selecting a subset of elements in the operator pool $\mathrm{T}$ to generate an ansatz operator. Operator selection is initialized with a starting ansatz operator which, for example, prepares the Hartree-Fock state for the given Hamiltonian. The selection then evaluates the commutator between each operator in the pool and the Hamiltonian with respect to the presently prepared state. The resulting vector of commutator evaluations defines a gradient that is steepest in the direction of the operator with the largest commutator. The operator corresponding to the largest commutator is then appended to the ansatz operator and, therefore, the ansatz circuit for the next evaluation of VQE. An associated parameter is introduced to tune the unitary rotation represented by the new operator, such that the number of optimized parameters grows with the size of the ansatz. When the norm of the gradient vector is less than a defined threshold, the algorithm converges and the lowest energy state is recovered. 

\subsection{Ansatz Compilation}
The above ansatz operators must be compiled into quantum circuits for execution. We use the XACC programming framework to synthesize circuits representing these ansatz operators. XACC provides a set of software methods to define and implement fermionic and spin transformations that reduce the operator expressions above into a sequence of discrete quantum gates \cite{mccaskey2020xacc}. In addition, the accompanying VQE and ADAPT-VQE methods orchestrate iterative execution of the ansatz and search through the corresponding parameter space. We defer details on the compilation methods to prior references \cite{mccaskey2019quantum}\cite{grimsley2019adaptive}. The results include, for example, the compiled one-parameter UCCD ansatz from \eqref{eq:yxxx} decomposed into the sequence of one- and two-qubit gate operations shown in Fig.~\ref{fig:UCC1}. Compiled circuits are then simulated and XACC facilitates this step as well. Here we perform numerical simulations that interface the compiled circuits with a simulator that otherwise faithfully execute the VQE and ADAPT-VQE methods. Details on the numerical simulations are reviewed below. 
\par 
We also study the effect of noise on circuits generated using the randomized compiling method that are logically equivalent to the UCCD one parameter ansatz but have a higher gate depth.  Specifically, we consider randomized compiling to transform ansatz circuits by introducing independent, random single-qubit gates such  that the output circuit is logically equivalent to the original \cite{RCcircuits}. The technique is implemented by dividing gate sets into so-called `easy' and `hard' gates and then reorganizing the original circuit into clock cycles alternating between a round of easy and hard gates applied to disjoint qubits per cycle \cite{RCcircuits}. Each round of easy gates is conjugated by a twirling gate $T_k$ from a twirling gate set and an inverse operator $T_{k-1}^c$. The original single qubit gates and the twirling gates are then compiled into new easy gate cycles. 
\par 

Randomized compiling was developed to address coherent and correlated noise by effectively transforming the noise into a stochastic Pauli noise channel. While randomized compiling can offer improvements in circuit performance \cite{hashim2021randomized}, our investigation does not test such promises due to the selected noise model described in the next section. Rather, we investigate the role that additional gates in the ansatz circuit have on the increased circuit depth and output state fidelity. We limit our analysis to randomized compiling of the UCCD ansatz  circuit due to its simplicity and single variational parameter.  At the time that we used the commercially available TrueQ software package to generate these circuits, it was not equipped to compile parametrized operators of the circuit.\cite{trueq}. 

\section{Numerical Methods}
\label{sec:Numerical Methods}
For our numerical studies, we use the model of a two-electron Hamiltonian representing sodium hydride (NaH) within the frozen-core approximation and with the STO-3G basis set. We require $n = 4$ qubits to encode the two spin-orbitals using the Jordan Wigner transformation, and the corresponding Pauli representation of terms in the electronic Hamiltonian is summarized in Table~\ref{tab:nah}. As described previously, we use the XACC framework to compile the ansatz circuits and perform  optimization of the ansatz parameters with respect to the estimated energy. The compiled circuits are simulated numerically using the IBM aer simulator as the XACC qpu backend, and we integrate a noise model that describes each gate by a noisy operation. 

\par In modeling noise in the quantum circuits, we assume a model by which each one- and two-qubit gates are followed by a depolarizing noise channel. The depolarizing noise model is a convenient device agnostic noise model that offers a coarse-grain representation for the loss of coherence caused by a noisy circuit, particularly for few-qubit numerical simulations \cite{IBMShortDepthCircs}. By setting the depolarizing noise to $p = 0$, we are able to simulate a noiseless circuit. We then chose a few parameters, where $p > 0$,  that mimic the settings of current hardware\cite{dahlhauser2021modeling}. Although it does not provide a fine-grain representation of the noise process, it has been used previously for accurately modeling the output from NISQ devices \cite{dahlhauser2021modeling}. In addition, randomized compiling is known to yield a statistical model for the quantum circuit noise that is well approximated by the depolarizing model and motivates our use for the model here \cite{ville2021leveraging}. 
\par
Within this model, the action of a single-qubit unitary gate $G$ acting on register element $j$ in the quantum state $\rho$ is simulated as
\begin{equation}
    \xi^{G}_{j}(\rho) = (1-p)\rho' + p \sum_{k} \sigma^{k}_{j} \rho' \sigma^{k}_{j}
\end{equation}
with $\rho' = G \rho G^{\dagger}$, $\sigma^{k}_{j} \in {X_j, Y_j, Z_j}$, and the noise parameter $p \in [0,1]$. Errors on a two-qubit gate acting on elements $i$ and $j$ are modeled similarly as $\xi^{G}_{i,j}(\rho) = \xi^{I}_{i}(\xi^{G}_{j}(\rho))$ with $I$ the identity operator. For our simulations, we use a gate set consisting of the single-qubit Pauli operator, the Hadamard gate, and single-qubit rotations with the two-qubit \textsc{cnot} gate. These gates are decomposed by XACC into the OpenQASM representation \cite{mckay2018qiskit}. Notably, OpenQASM gates that are diagonal in the computational basis, e..g, $Z$, are not modeled with noise, a consequence of using the IBM aer simulator. For our noise model, we use noise levels for the two-qubit \textsc{cnot} gate that is always ten times the value of the single-qubit gate noise. This is motivated by the difference in noise values observed in hardware. We use the same noise level for all single-qubit gates. 
\par
For different values of noise, we explore the influence of methodology on parameter optimization by comparing results from two different optimizers, COBYLA and L-BFGS. These two optimizers were chosen as gradient-free and gradient-based methods respectively. Constrained optimization by linear approximation (COBYLA)  constructs and optimizes a series of linear approximations to the objective function to compute parameter steps that minimize the objective function \cite{cobyla}. We use the third-party library implementation nlopt with a maximum number of iterations set to 1,000 with a convergence tolerance of $10^{-6}$ \cite{nlopt}. By comparison, the low-memory Broyden-Fletcher-Goldfarb-Shanno (L-BFGS) method is a quasi-Newton method that uses the gradient and an approximate Hessian to determine steps in parameter space that minimize the objective function \cite{lbfgs,lbfgs2}. We use the third-party library implementation mlpack with the central gradient method with a maximum number of iterations set to 500,000 with a convergence tolerance of $10^{-4}$ and numerical step set to $10^{-7}$ \cite{mlpack}.

\par 
The output of each simulation is a list of expectation values characterizing the individual Pauli terms representing the Hamiltonian shown in Table~\ref{tab:nah}, which are then combined to estimate the electronic energy. In addition, the output includes the simulated quantum state in the form of a density matrix. We compare the simulated energy estimate and simulated state with the energy and state obtained for the same Hamiltonian molecule calculated using the complete active-space self-consistent field (CASSCF) method within pyscf \cite{sun2018pyscf}. We investigate different levels of noise as well as the behavior with respect to inter-nuclear configuration.

\begin{table}[H]
\centering
\caption{Four-spin representation of the two-electron, NaH Hamiltonian and expectation value of the Pauli strings with respect to the UCCD ansatz for $r = 1.91438 \text{{\AA}}$.}
\label{tab:nah}
\begin{tabular}{|l|r|r|r|}
\hline
$j$ & $c_j(r_0)$ [Ha] & $P_j$ & $\langle P_j(\theta) \rangle$ \\ \hline
0 & -159.40289 & $I_0 I_1 I_2 I_3$ &  1\\ \hline
1 & 0.0323625 & $X_0 X_1 I_2 I_3$ & 0 \\ \hline
2 & 0.0202421 & $X_0 X_1 X_2 X_3$ & $-\sin\theta$ \\ \hline
3 & 0.0202421 & $X_0 X_1 Y_2 Y_3$ & $-\sin\theta$  \\ \hline
4 & 0.0229208 & $X_0 X_1 Z_2 I_3$ &  0\\ \hline
5 & -0.00944179 & $X_0 X_1 I_2 Z_3$ & 0 \\ \hline
6 & 0.0323625 & $I_0 I_1 X_2 X_3$ &  0\\ \hline
7 & 0.0323625 & $Y_0 Y_1 I_2 I_3$ &  0\\ \hline
8 & 0.0202421 & $Y_0 Y_1 X_2 X_3$ & $-\sin\theta$ \\ \hline
9 & 0.0202421 & $Y_0 Y_1 Y_2 Y_3$ & $-\sin\theta$  \\ \hline
10 & 0.0229208 & $Y_0 Y_1 Z_2 I_3$ & 0 \\ \hline
11 & -0.00944179 & $Y_0 Y_1 I_2 Z_3$ & 0 \\ \hline
12 & 0.0323625 & $I_0 I_1 Y_2 Y_3$ & 0 \\ \hline
13 & 0.0149385 & $Z_0 I_1 I_2 I_3$ & $-\cos\theta$ \\ \hline
14 & 0.0229208 & $Z_0 I_1 X_2 X_3$ & 0 \\ \hline
15 & 0.0229208 & $Z_0 I_1 Y_2 Y_3$ & 0 \\ \hline
16 & 0.0816923 & $Z_0 Z_1 I_2 I_3$ & $-1$ \\ \hline
17 & 0.158901 & $Z_0 I_1 Z_2 I_3$ & 1 \\ \hline
18 & 0.101934 & $Z_0 I_1 I_2 Z_3$ & $-1$ \\ \hline
19 & -0.387818 & $I_0 Z_1 I_2 I_3$ &  $-\cos\theta$ \\ \hline
20 & -0.00944179 & $I_0 Z_1 X_2 X_3$ & 0 \\ \hline
21 & -0.00944179 & $I_0 Z_1 Y_2 Y_3$ &  0\\ \hline
22 & 0.101934 & $I_0 Z_1 Z_2 I_3$ & $-1$ \\ \hline
23 & 0.117450 & $I_0 Z_1 I_2 Z_3$ & 1 \\ \hline
24 & 0.0149385 & $I_0 I_1 Z_2 I_3$ & $-\cos\theta$ \\ \hline
25 & 0.0816923 & $I_0 I_1 Z_2 Z_3$ & $-1$ \\ \hline
26 & -0.387818 & $I_0 I_1 I_2 Z_3$ & $\cos{\theta}$ \\ \hline
\end{tabular}
\end{table}

\section{Results}
\label{sec:results}
\par
We conduct a range of studies of the above noise model applied to VQE and ADAPT-VQE simulations of the NaH molecule. We compare the statistics and fidelity of the expectation values of the noisy and noiseless simulations across a range of values for the internuclear distance, $R$, and different optimizers.  For the studies with VQE we study the simulations of the original circuit and compare them with the results of the randomly compiled circuits. The goal is to understand how noise affects these properties and if there are parameters that are more resilient to noise than others.

\begin{figure}
    \centering
    \includegraphics[scale=0.4]{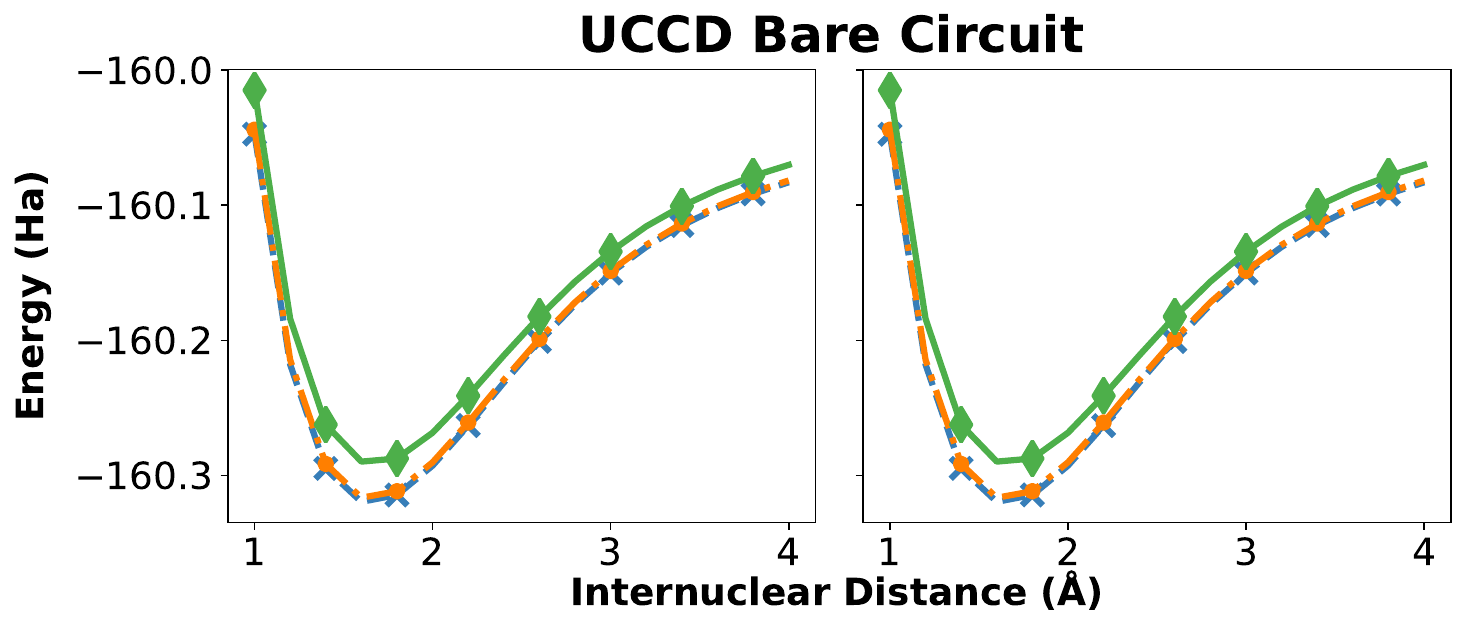}
    \includegraphics[scale=0.4]{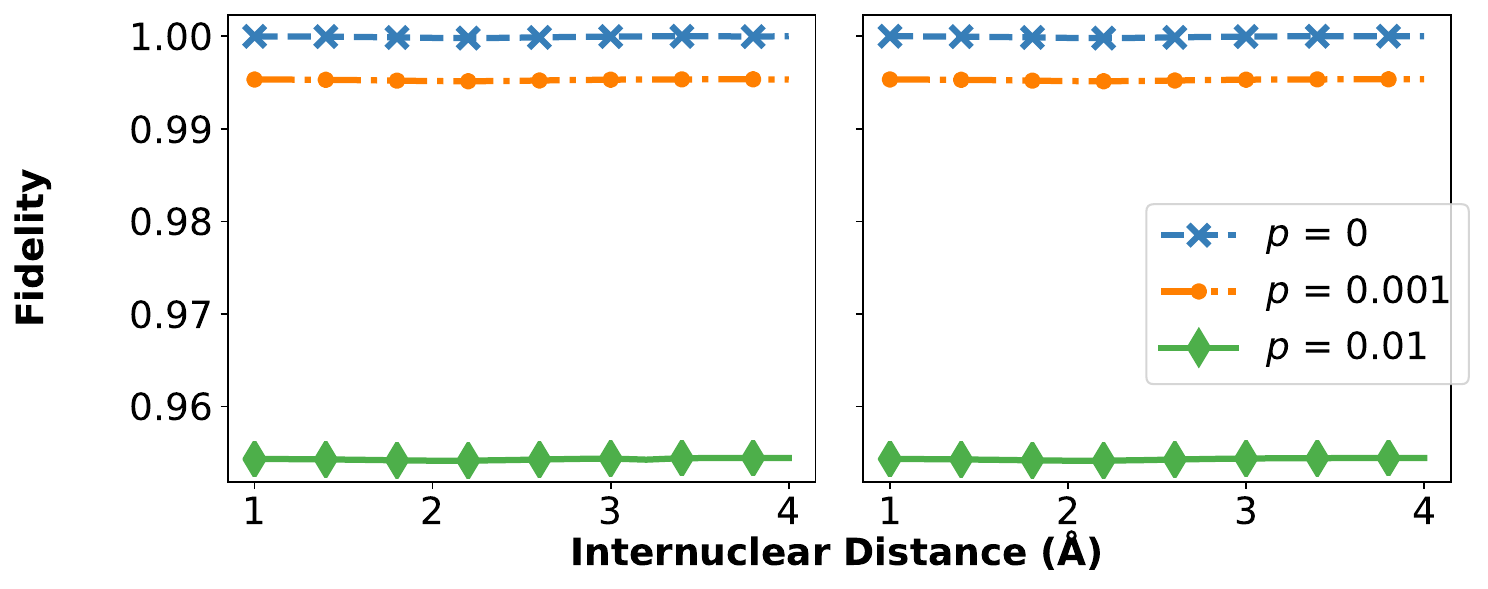}
    \caption{The top row presents the ground state electronic energy of NaH using the UCCD bare circuit ansatz within the VQE method with different levels of depolarizing noise, $p$. The bottom row presents the corresponding fidelity of the prepared quantum state and that of the ground state found through conventional means. The left and right columns differentiate the results found using the COBYLA and L-BFGS optimizers respectively.}
    \label{fig:BCVQE}
\end{figure}

\begin{figure}[!ht]
    \centering
    \includegraphics[scale=0.5]{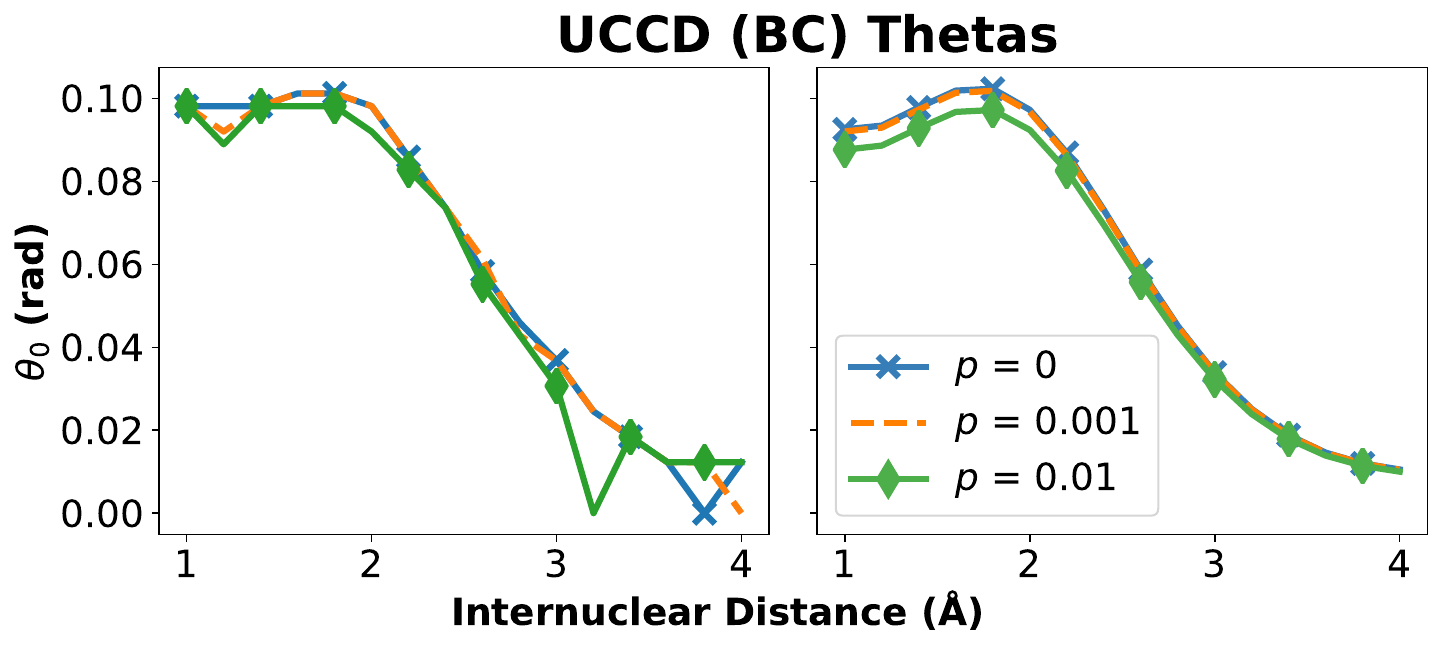}
    \caption{The optimal parameters for the ground state energy of the bare UCCD ansatz found using the Cobyla (left) and L-Bfgs (right) optimizers.}
    \label{fig:BCThetas}
\end{figure}
\par 
We begin our study of the results of the ground state electronic energy calculation and fidelity of the prepared state with our simplest ansatz, the UCCD bare circuit (BC) ansatz, shown in figure~\ref{fig:BCVQE}. The simulated ground state energy predictably increases with increases in the depolarizing noise parameter $p$ across all $R$. In addition the fideitly accross all $R$ drops as we increase the depolarizing parameter. The increase in energy from the noisless simulation is no greater than 0.03 Ha at the greatest level of noise, while the fidelity drops less than 5\%. Additionally, the overall shape of the curve is fairly standard for a potential energy curve, and the fidelity is relatively flat across all $R$. We report little difference between the optimizers used to find the energy and the fidelity of the this ansatz. Figure~\ref{fig:BCThetas} plots the optimal parameters found to represent the ground state in the UCCD ansatz using the COBYLA and L-BFGS optimizers. There are noticeable differences in the parameters returned by COBYLA and L-BFGS with the latter showing a smoother variation across different levels of the noise parameter, especially at larger values of $R$. These differences however, didn't correspond to any significant difference to the reported energy or fidelity of the respective optimizer. 

\begin{figure}[!ht]
    \centering
    \includegraphics[scale=0.4]{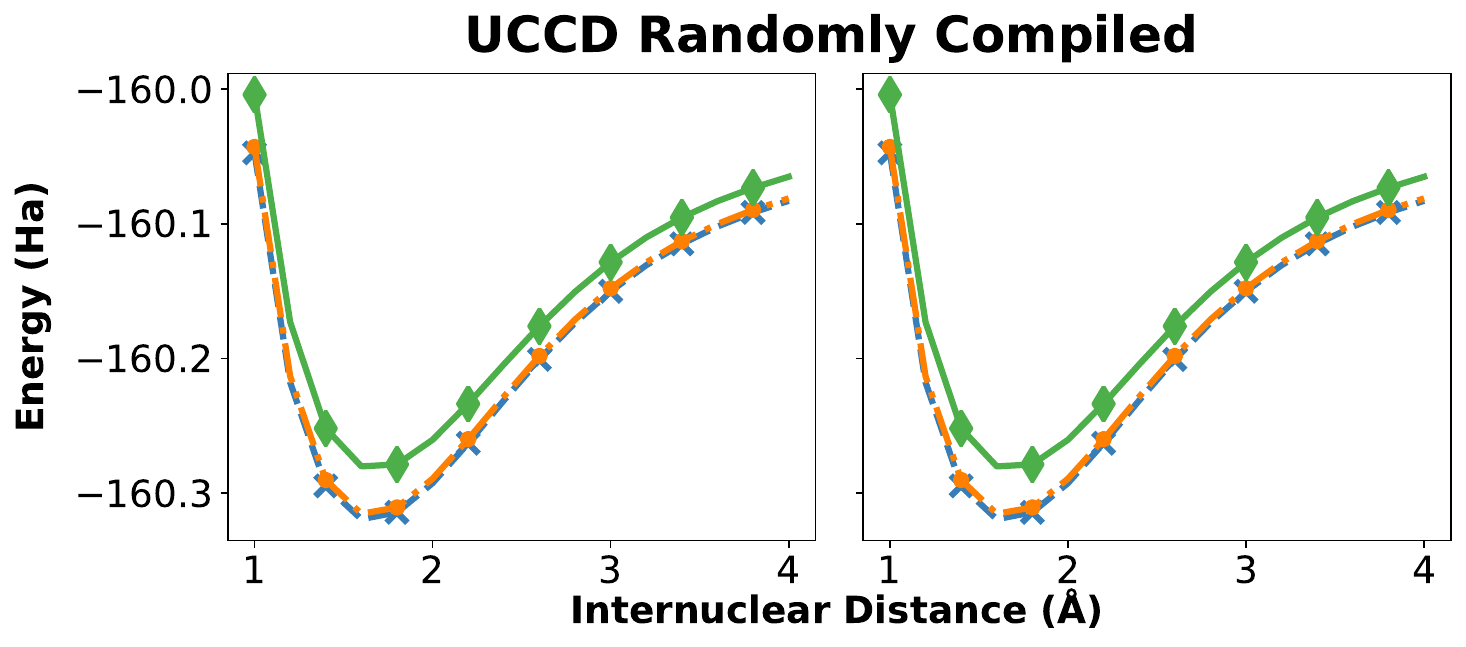}
    \includegraphics[scale=0.4]{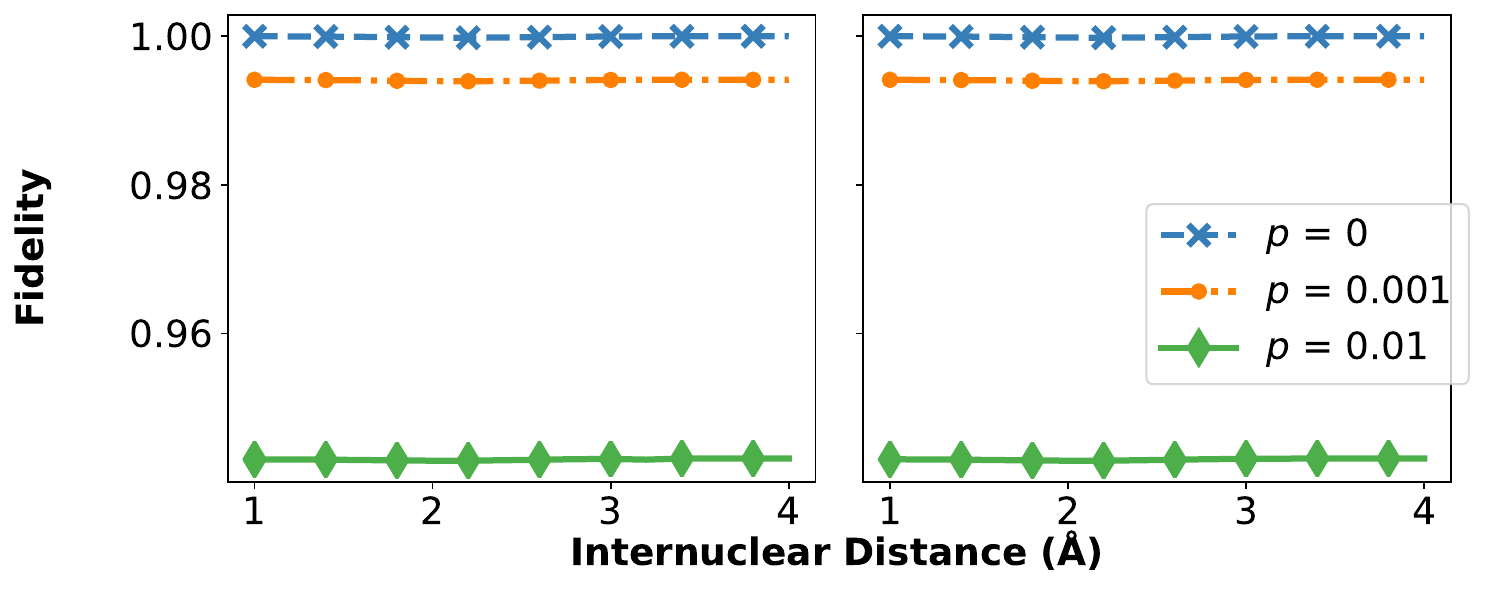}
    \caption{The top row presents the ground state electronic energy of NaH using the UCCD randomly circuit ansatz within the VQE method with different levels of depolarizing noise, $p$. The bottom row presents the corresponding fidelity of the prepared quantum state and that of the ground state found through conventional means. The left and right columns differentiate the results found using the COBYLA and L-BFGS optimizers respectively.}
    \label{fig:RCVQE}
\end{figure}

\begin{figure}[!ht]
    \centering
    \includegraphics[scale=0.5]{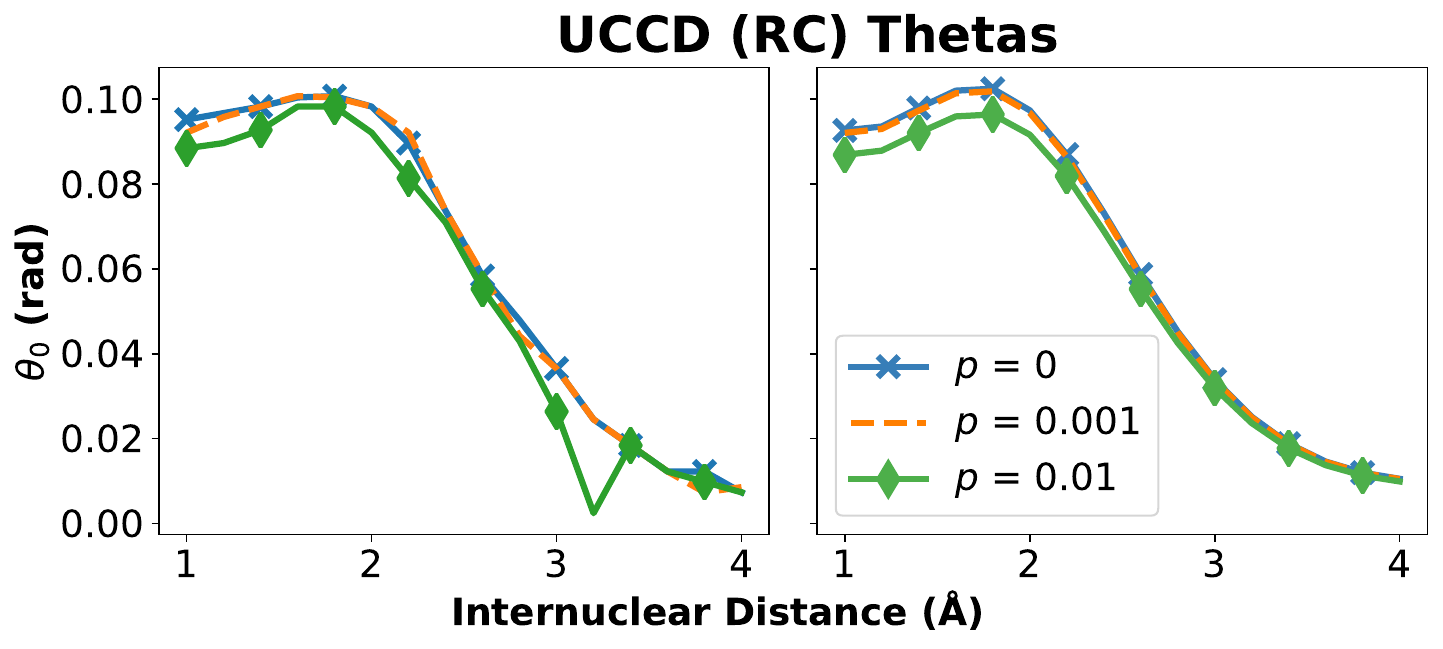}
    \caption{The optimal paramters for the ground state enegy of the randomly compiled UCCD ansatz found using the Cobyla (left) and L-Bfgs (right) optimizers.}
    \label{fig:RCThetas}
\end{figure}
\par
Figure~\ref{fig:RCVQE} plots the results of the UCCD randomly compiled (RC) ansatz simulation at varying levels of depolarizing noise. Similar to the bare UCCD ansatz, the simulated ground state energy increases with increases in the noise parameter across all $R$. Again, we see a corresponding drop in fidelity as we increase the noise parameter, and minimal differences in the reported energy and fidelity when comparing the COBYLA and L-BFGS optimizers. The increase in energy from the noiseless simulation is no greater than 0.04 Ha at the greatest level of noise, while the fidelity drops less than 6\%.The overall shape of the curve is standard for a potential energy curve, and the fidelity is relatively flat across all $R$.The plotted thetas in figure~\ref{fig:RCThetas} show a different selection of optimal parameters between the COBYLA and L-BFGS optimizers. Like the bare UCCD ansatz,there are obvious difference between the selected optimal parameter for the ground state as the depolarizing noise increases for both optimizers. For example, the COBYLA optimizer has a greater varriance than that of the L-BFGS optimzer, especially at higher values of $R$. When comparing the differences in optimal parameter selection between optimizers at the same level of noise, we report no significant difference to the resultant energy or fidelity between the respective optimizers.

\begin{figure}[!ht]
    \centering
    \includegraphics[scale=0.4]{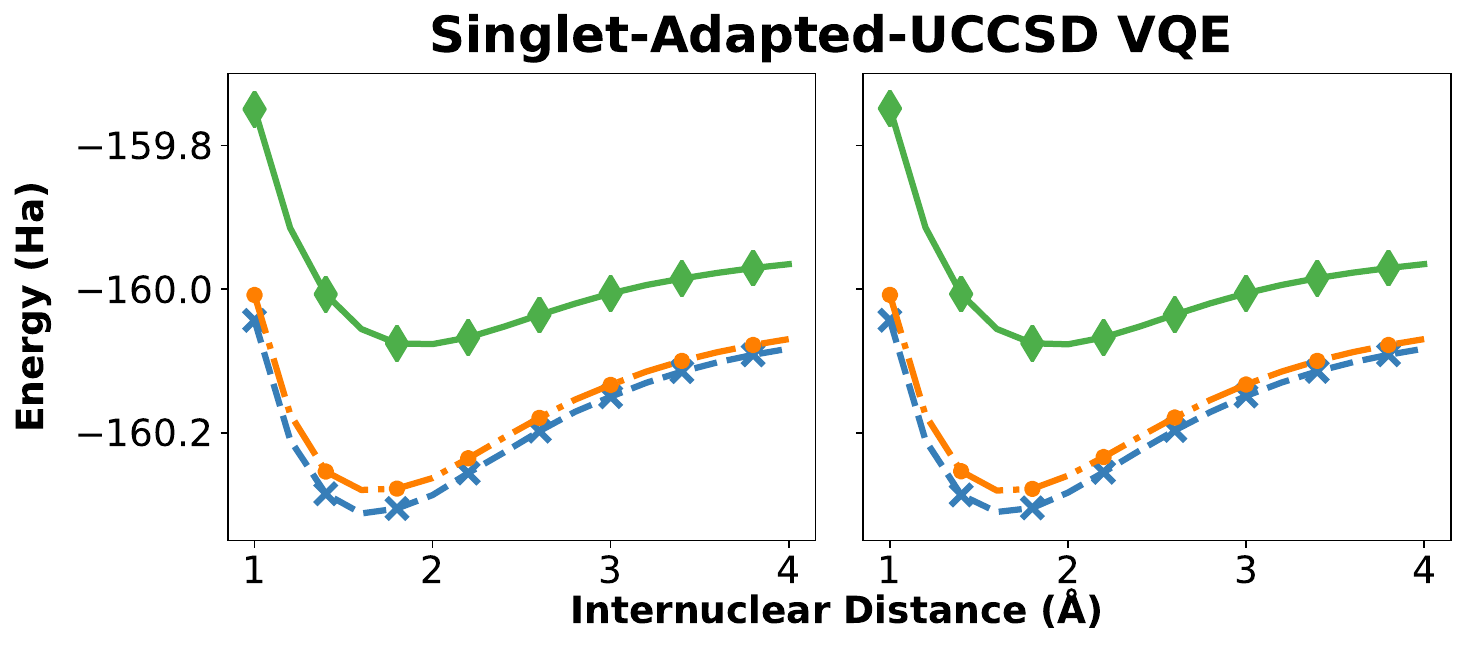}
    \includegraphics[scale=0.4]{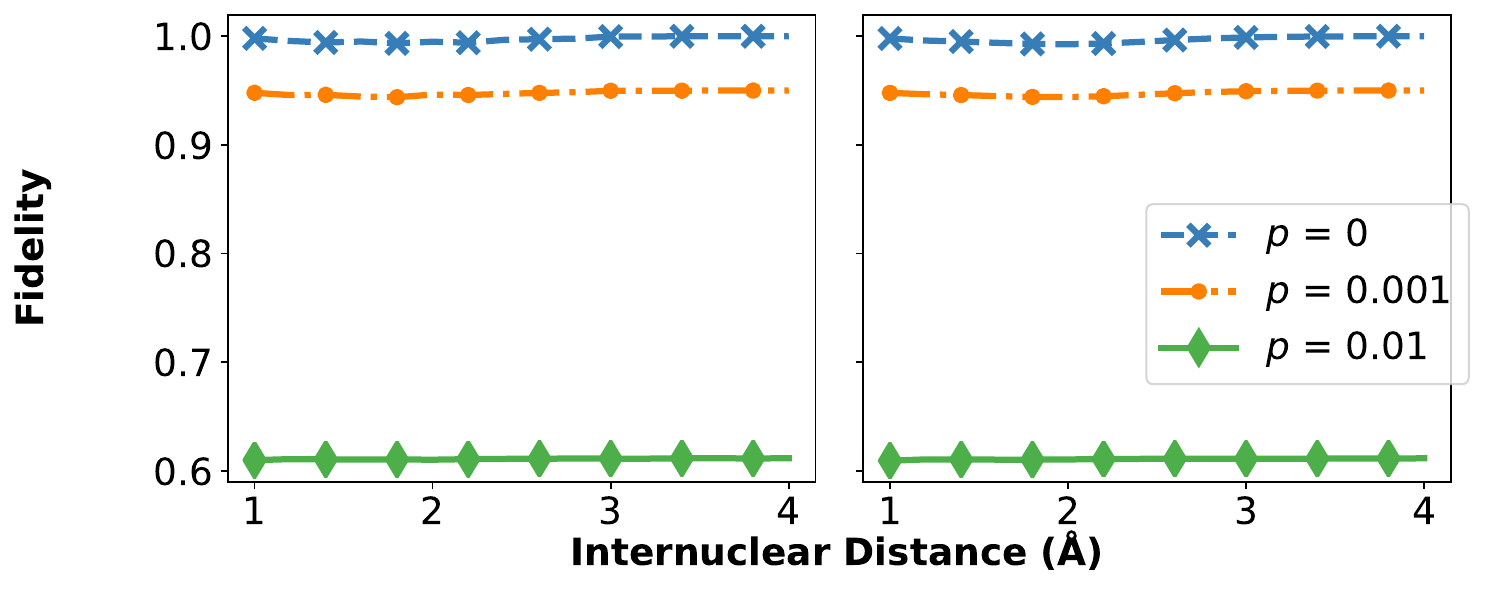}
    \caption{The top row presents the ground state electronic energy of NaH using the Singlet-Adapated-UCCSD ansatz within the VQE method with different levels of depolarizing noise, $p$. The bottom row presents the corresponding fidelity of the prepared quantum state and that of the ground state found through conventional means. The left and right columns differentiate the results found using the COBYLA and L-BFGS optimizers respectively.}
    \label{fig:SAVQE}
\end{figure}

\begin{figure}[!ht]
    \centering
    \includegraphics[scale = 0.45]{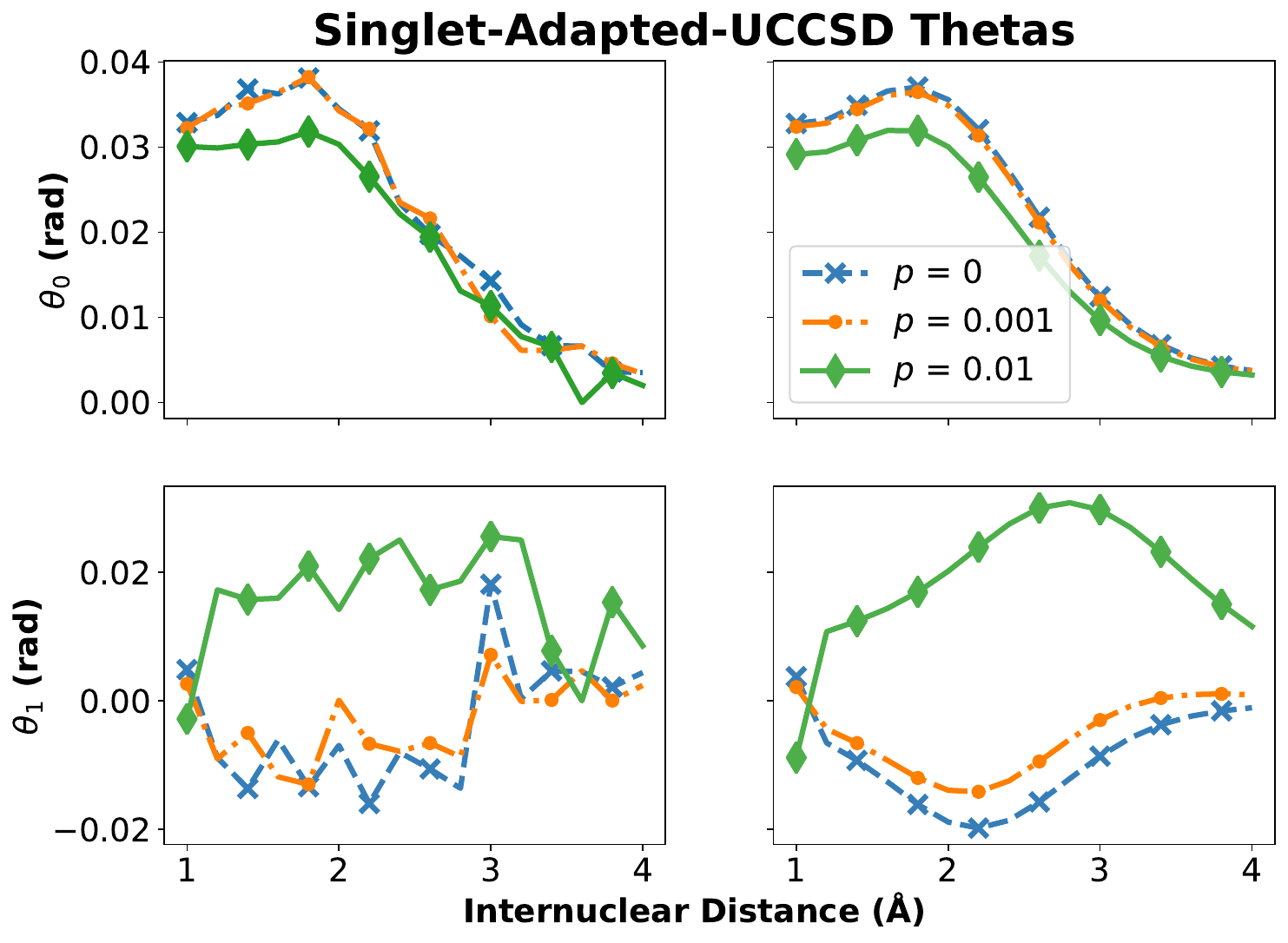}
    \caption{The optimal values of the circuit parameters (top row) $\theta_0$ and (bottom row) $\theta_1$ for the singlet-adapted UCCSD ansatz as a function of obtained from VQE with respect to the NaH internuclear distance when  using the COBYLA (left column) and L-BFGS (right colum) optimization methods.}
    \label{fig:SA_Thetas}
\end{figure}

\par
The energy and fidelity results of the Singlet-Adapted-UCCSD ansatz are reported in figure~\ref{fig:SAVQE}. The general trend of increasing energy and decreasing fidelity as the depolarizing noise increases is still seen within this plot. The general shape of the energy curve is standard to that of potential energy curves, and the fidelities are relatively flat across all $R$. While there is negligible difference between the reported values of the energy and fidelity between optimizers, there is significant difference between the reported values at varying levels of the noise parameter. The increase in energy from the noiseless simulation is no greater than 0.3 Ha, while the fidelity drops less than 40\% at the greatest value of noise. Figure~\ref{fig:SA_Thetas} show a different selection of optimal parameters between the COBYLA and L-BFGS optimizers. Both optimizers search similar regions but there is noticeable variation in their results. In addition, as the noise level $p$ increases, the parameter search changes in tandem, with the $p$ = 0.01 case for the $\theta_1$ parameter showing the most dramatic shift. In general,the COBYLA optimzer varies greatly with a much more jagged curve as we search across all $R$ for both $\theta_0$ and $\theta_1$ when we compare it to that of the L-BFGS optimizer. While being smoother, the L-BFGS optimizer reports similar values of the energy and fidelity to that of the COBYLA optimizer. 

\begin{figure}
    \centering
    \includegraphics[scale=0.4]{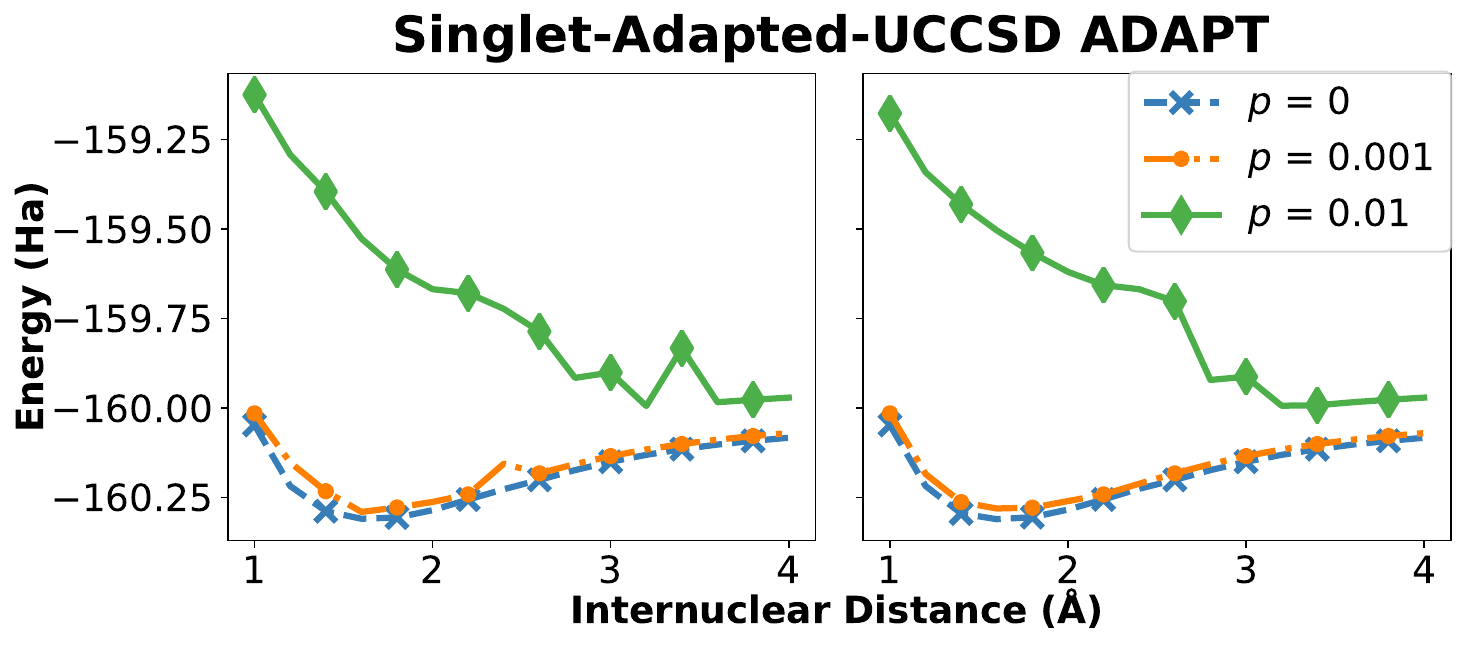}
    \includegraphics[scale=0.4]{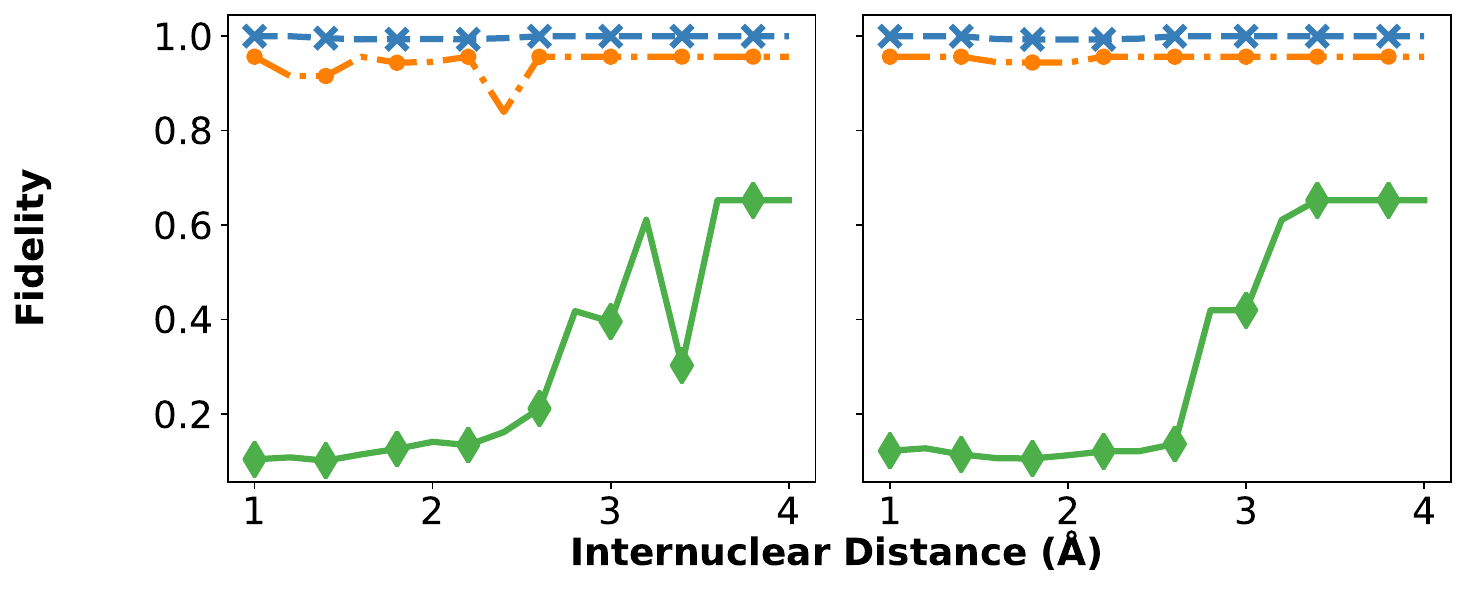}
    \caption{The top row presents the ground-state electronic energy of NaH calculated using ADAPT-VQE with Singlet-Adapted-UCCSD operators while the bottom row presents the corresponding fidelity of the prepared quantum state with the ground state calculated using conventional methods. The left and right columns distinguish between the parameters chosen through the COBYLA and L-BFGS optimization routines respectively.}
    \label{fig:AVQE}
\end{figure}

\begin{figure}[]
    \centering
    \includegraphics[scale = 0.4]{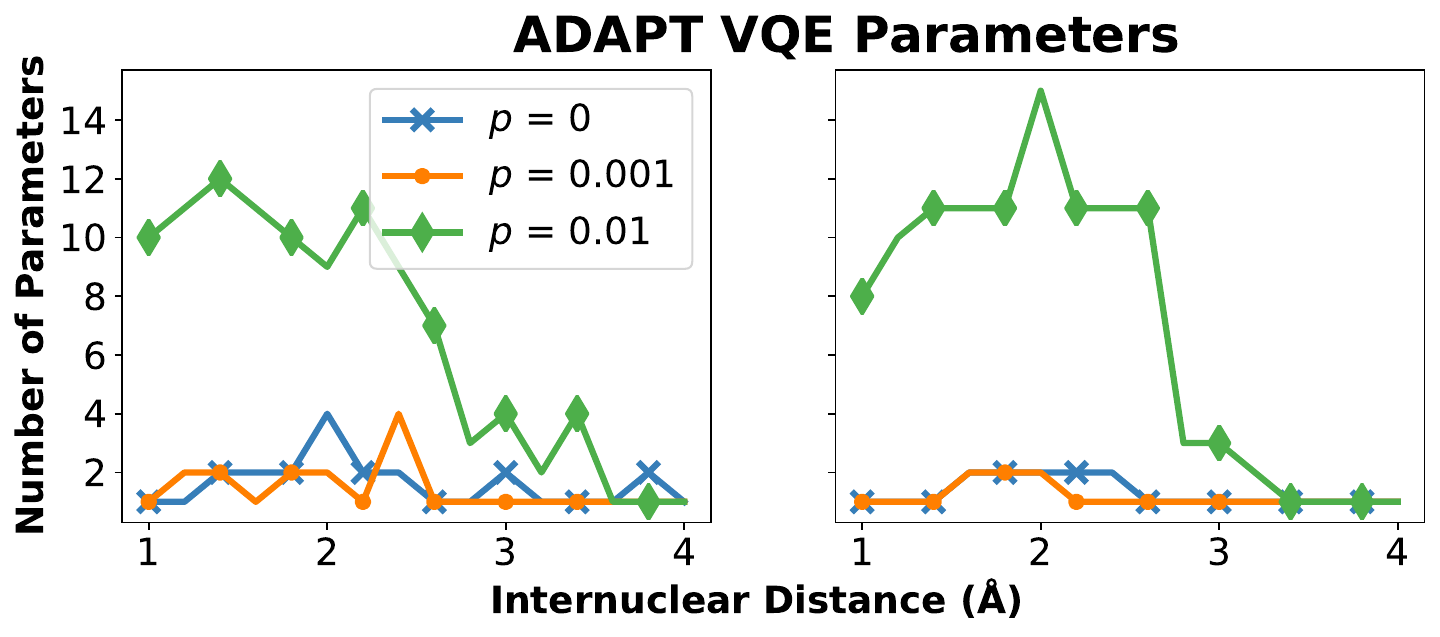}
    \caption{The number of optimal parameters chosen by the ADAPT algorithm to reach convergence as a function of the internuclear distance, using the COBYLA (left) and L-BFGS (right) optimizers.}
    \label{fig:Adapt_thetas}
\end{figure}

\par
The energy and fidelity results of the ADAPT VQE simulation using the singlet-adapted-uccsd operator pool is shown in figure~\ref{fig:AVQE}. The general trend of increasing energy and decreasing fidelity is still seen within in this plot, however with increasing depolarizing noise, the shape of the energy plot deviates from standard potential enegy curves. Likewise, the fidelity of the the state prepared is not flat across all $R$. The increase in energy from the noiseless simulation is no greater than 1.0 Ha, while the fidelity drops at most a little less than 90\% at the greatest value of noise. Contrast to the previous ansatze, the ADAPT-VQE ansatze grow uniquely with respect to $R$. This growth can be shown by plotting the number of parameters the ADAPT-VQE algorithm used in ansatz construction to reach convergence. which is shown in figure~\ref{fig:Adapt_thetas}. There are significant differences in the number of parameters between the COBYLA and L-BFGS optimizers,however,the general trend in this plot shows how the number of parameters decrease as $R$ increases.
\par

\begin{figure}
    \centering
    \includegraphics[scale=0.31]{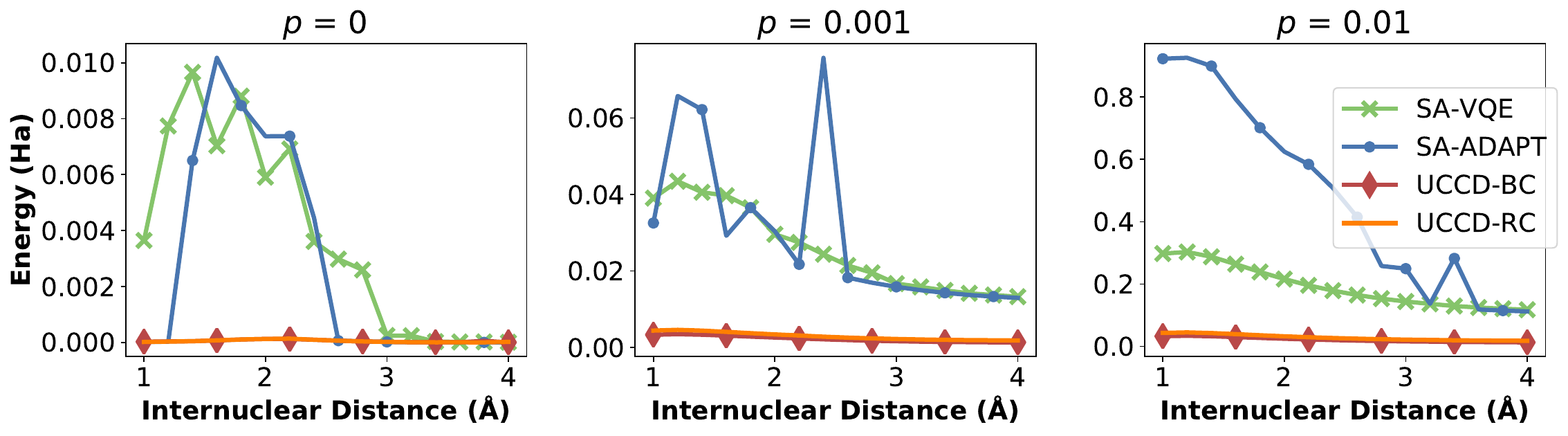}
    \caption{The absolute error between the electronic energy of the simulated ansatze and the ground truth found through conventional means. All results shown use the COBYLA optimizer.}
    \label{fig:ERROR}
\end{figure}

\par
Figure~\ref{fig:ERROR} plots the absolute error in energy of all the above ansatze simulated from the ground state enegy found through conventional means. As we presented the energy and fidelity data of the different ansatze from the smallest to largest circuits in terms of gate depth, we see a similar trend in the absolute error plot. This is because every one- and two-qubit gate in every circuit is affected by depolarizing noise, thus increasing the number of gates increases the amount of noise in the system. The bare UCCD ansatz is the smallest ansatz and as such sees the smallest change in energy at the greatest value of depolarizing noise. Similarly, the randomly compiled circuits result in slightly larger increase for the same noise level due to the presence of additional one-qubit gates. On average, the randomly compiled circuits consist of approximately 50\% more single-qubit gates than the bare UCCD ansatz. The Singlet-Adapted-UCCSD ansatz has more gates, but more significantly, the ansatz has a higher number of two-qubit, \textsc{CNOT}, gates than that of the previous ansatze. The effect of depolarizing noise is an order of magnitude greater within two-qubit gates than that of one-qubit gates. This translates to a greater increase in energy as the depolarizing parameter increases. Lastly, the ADAPT-VQE ansatze exhibit the greatest increase in energy due to the large increase in one- and two-qubit gates within the circuit. The number of one- and two-qubit gates required for $p=0.01$ is approximately an order of magnitude higher than those found for any other ansatze tested.

\section{Conclusions}
\label{sec:conc}
Our goal was to understand how noise affects the energy and fidelity of the ansatz and if there are parameters that are more resilient to noise than others. Across all ansatz circuits, we find that the noiseless simulations always recover the ideal energy and unit fidelity, while in the presence of noise, the depth of the ansatz circuit has the most important influence on the error in these quantities. Increases in gate depth of the ansatz necessarily increases the error in energy and lowers the fidelity and this is amplified at larger noise values. The differences observed between the bare UCCD and the randomly compiled UCCD circuits reflect differences in their single-qubit gate depth while the Singlet-Adapted-UCCSD and ADAPT-VQE related ansatze, reflect differences in the number of two-qubit gates. More specifically, this indicates that the number of \textsc{cnot} gates is the leading limit on how accurately the ansatz can describe the ground-state wavefunction.
\par
The COBYLA and L-BFGS optimizers do yield different optimal parameters for the UCCD bare circuit, the UCCD randomly compiled circuit, and the Singlet-Aapted-UCCSD circuti however, these differences generally have a small influence on the outcomes of the fidelity and energy. For example, prior work has confirmed that the energy surface characterizing the UCCD ansatz is relatively smooth \cite{NoisyVQEIEEE}, while our results presented here confirm similar behavior in the presence of noise. The differences between the two optimizer are more apparent within the ADAPT-VQE method at higher levels of noise, but generally follow the same t
\par 
 A leading feature of the ADAPT-VQE algorithm is that the ansatz grows iteratively to improve the accuracy of the computed energy. This feature allows for variability in the number of and types of gates that make up the ansatz. However, as shown here, variability in the ansatz circuit depth due to increasing depolarizing noise, leads to fluctuations in the energy and fidelity that are not observed within the fixed ansatz. 
\par 
We studied how depolarizing noise specifically affected different ansatze, but note that other sources of noise could be present in other systems. Thus, this study would not accurately represent systems that included more, or different types of noise such as readout noise or coherent noise. In addition, we are limited to the size of the systems we are able to study due to the use of density matrices which are exponentially larger than statevectors. Finally, we note that the ADAPT-VQE method could be further modified to combat the effects of noise such as modifying the ADAPT-VQE configuration to ensure continuity in the ansatz circuit depth and parameter values. For example, intelligent pruning of the ADAPT ansatz may be beneficial with little loss of accuracy. We defer this point for later work. 

\section{Declarations}
\begin{backmatter}

\section*{Availability of data and materials}
The data and materials are available upon request. 
\section*{Acknowledgements}
This research used resources of the Oak Ridge Leadership Computing Facility, which is a DOE Office of Science User Facilities supported by the Oak Ridge National Laboratory under Contract DE-AC05-00OR22725. This research used resources of the Compute and Data Environment for Science (CADES) at the Oak Ridge National Laboratory, which is supported by the Office of Science of the U.S. Department of Energy under Contract No. DE-AC05-00OR22725. 

\section*{Funding}
This work was supported by the “Embedding Quantum Computing into Many-body Frameworks for Strongly Correlated Molecular and Materials Systems” project, which is funded by the U.S. Department of Energy (DOE), Office of Science, Office of Basic Energy Sciences, the Division of Chemical Sciences, Geosciences, and Biosciences. 
This work was also supported by the Quantum Science Center (QSC), a National Quantum Information Science Research Center of the U.S. Department of Energy (DOE).

\section*{Competing interests}
The authors declare that they have no competing interests.

\section*{Authors' contributions}
J.~W.~and M.~G.~developed the methods, performed the experiment, processed the data, and composed the manuscript. D.C.~developed the methods and composed the manuscript, P.C.L.~processed the data and composed the manuscript, T.N.~developed the methods, A.J.M.~developed the methods, and T.S.H.~processed the data and composed the manuscript.

\bibliographystyle{bmc-mathphys} 
\bibliography{references}      


\end{backmatter}
\end{document}